\documentclass[twocolumn,prl,showpacs]{revtex4}%
\usepackage{graphicx}%
\usepackage{amsmath}%
\usepackage{amsfonts}%
\usepackage{amssymb}

\def\d{{\partial}}
\def\s{{\sigma}}
\def\e{{\epsilon}}
\def\k{{ {\bf k} }}

\def\w{{\omega}}
\def\a{{\alpha}}

\def\g{{\gamma}}

\begin{document}
\title{Intrinsic Spin Hall Effect in $s$-wave Superconducting State: \\
Analysis of Rashba Model}
\author{H. \textsc{Kontani}, J. \textsc{Goryo} and D.S. \textsc{Hirashima}}
\date{\today }

\address{
Department of Physics, Nagoya University,
Furo-cho, Nagoya 464-8602, Japan. 
}

\begin{abstract}
A general expression for the spin Hall conductivity (SHC) 
in the $s$-wave superconducting state at finite temperatures is derived.
Based on the expression, 
we study the SHC in a two-dimensional electron gas model
in the presence of Rashba spin-orbit interaction (SOI).
The SHC is zero in the normal state, whereas it takes a 
large negative value as soon as the superconductivity occurs,
due to the change in the quasiparticle contributions.
Since this remarkable behavior is independent of 
the strength of the SOI, it will be widely observed 
in thin films of superconductors with surface-induced Rashba SOI,
or in various non-centrosymmetric superconductors.
\end{abstract}

\pacs{72.25.Ba,74.70.-b,72.10.-d}
\maketitle


\sloppy

The spin Hall effect (SHE), which is a phenomenon
that an electric field ${\bf E}$ induces a transverse
spin current ${\bf J}^{\rm S}$,
has attracted considerable attention.
The intrinsic SHE, which is independent of impurity scattering,
was first predicted in semiconductors 
\cite{Sinova-SHE,Murakami-SHE,Inoue-SHE,Nomura-SHE}.
Later, it was revealed that the large intrinsic SHE emerges 
in transition metals and their compounds, since the
complex $d$-orbital wavefunction induced by spin-orbit interaction (SOI)
gives rise to the ``{orbital Berry phase}'' that works as
a spin-dependent effective magnetic field
 \cite{Kontani-Ru,Kontani-Pt,Guo-Pt,Tanaka-4d5d}.
In several metals,
the spin Hall conductivity (SHC, $\s_{\rm SH}$) had been experimentally 
determined by observing the inverse SHE signal
\cite{Saitoh,Kimura,Valenzuela}.
The observed SHC in Pt exceeds 
$200\ \hbar e^{-1}\cdot \Omega ^{-1}\mathrm{cm}^{-1}$ \cite{Kimura},
and the SHCs in Nb and Mo are negative \cite{Otani-Nb}. 
These results can be explained in terms of the intrinsic SHE
 \cite{Tanaka-4d5d,OHE},
suggesting the importance of the intrinsic mechanism in transition metals.

In spintronics, superconductivity is widely used to control the spin state.
In this sense, a natural question is whether and how the SHE emerges
in the superconducting state.
Although DC electric field ${\bf E}$ cannot exists in bulk superconductors,
the SHC in superconductors is observable as follows:
For example, the Hall voltage due to inverse SHE 
should appear in the superconducting tunneling junction, 
thanks to the charge imbalance effect \cite{Takahashi}.
Moreover, the gradient of temperature will induces the transverse 
spin current in a bulk superconductors when the SHC is finite:
The observed spin Nernst conductivity
$\a_{\rm SH} \equiv J_y^{\rm S}/(-\nabla_x T)$
 \cite{spinNernst}
would be related to the SHC by the Mott relation
$\s_{\rm SH} \propto T \s_{\rm SH}'(\e_{\rm F})$,
as discussed in the intrinsic anomalous Hall effect \cite{Niu06}.
In addition, 
the SHC in type-II superconductors will be measurable in the mixed state 
under the magnetic field, since the resistivity is nonzero \cite{Saitoh-p}.
Therein, SHC will not be affected by moving vortices
since they do not convey spin.

In this letter, we present a general expression for the intrinsic SHC 
in the superconducting state.
This expression indicates that the current vertex correction (CVC), 
which is a consequence of the conservation laws in the field theory,
can cause a significant change in the SHC in the superconducting state.
Based on the expression,
we analyze the two-dimensional electron gas (2DEG) model with Rashba SOI
as a typical spin Hall system, 
and reveal that a giant SHC emerges in the superconducting state
due to the CVC, independently of the strength of the SOI.
This phenomenon will be observed in thin film superconductors
with the aid of surface-induced Rashba SOI \cite{surface}, 
or in various non-centrosymmetric superconductors
such as CePt$_3$Si, Li$_2$Pt$_3$B, MgSi, and Y$_2$C$_3$.

The Rashba 2DEG model
describes the electrons in a semiconductor inversion layer.
It had been studied intensively not only as the issue of 
semiconductor spintronics, 
but also as a model for non-centrosymmetric superconductor \cite{Sigrist}.
The SHC in Rashba 2DEG model
was first studied by Sinova et al \cite{Sinova-SHE}.
They found that the SHC takes a universal value $-e/8\pi$
in the ballistic regime.
($-e$ is the electron charge.)
Later, Inoue et al \cite{Inoue-SHE} had shown that the SHC vanishes identically
in the diffusive regime due to the CVC induced by short-range impurities.
This fact is also explained by the relation
${\dot {\hat S}}_x\equiv i[{\hat H},{\hat S}_x]\propto {\hat J}_x^{\rm S}$
in the Rashba model \cite{Nomura-SHE,Dimitrova}, where 
${\hat J}_x^{\rm S}$ is the spin current.
However, this relation does not hold in the superconducting state.
For this reason, the SHC in Rashba 2DEG model 
shows a {\it considerably large value just below $T_{\rm c}$
unless the SOI is zero}, as we will show below.

The $s$-wave BCS-Rashba 2DEG model is expressed 
by the following $4\times4$ form in the Nambu representation
 \cite{Sigrist}:
\begin{eqnarray}
{\hat H} = \frac12 \sum_\k {\hat \phi}_\k^\dagger 
{\hat H}_\k {\hat \phi}_\k, \ \ 
{\hat H}_\k=
\left(
\begin{array}[c]{cc}
{\hat h}_\k^0 & -i\Delta{\hat \s_y} \\
i\Delta{\hat \s_y} & -{\hat h}_{-\k}^{0*}
\end{array}
\right) ,
 \label{eqn:Hk}
\end{eqnarray}
where ${\hat \s}_i$ ($i=x,y,z$) is the Pauli matrix, and
${\hat \phi}_\k^\dagger=(c_{\k\uparrow}^\dagger,c_{\k\downarrow}^\dagger,
c_{-\k\uparrow},c_{-\k\downarrow})$;
$c_{\k\s}$ is an annihilation operator of electron.
${\hat h}_\k^0=\e_\k {\hat 1}+\lambda({\hat \s}_x k_y - {\hat \s}_y k_x)$
is the $2\times2$ $k$-linear Rashba 2DEG Hamiltonian in the normal state,
where $\lambda$ is the SOI parameter and 
$\e_\k=\k^2/2m-\mu$; $\mu$ is the chemical potential.
$\Delta$ is the superconducting gap;
we assume that $0\le \Delta\ll \mu$ hereafter.
The quasiparticle spectrum is given by 
$E_{\k,\pm}= \sqrt{(\e_\k\pm\lambda k)^2+\Delta^2}$.
We also we assume that the spin-orbit splitting,
$\Delta_{\rm SO}= 2\lambda k_{\rm F}$,
is much smaller than $\mu$.
In this case, the triplet pairing
induced by the Rashba SOI is negligible \cite{Sigrist}.

Below, we study the SHC in the presence of 
short-range nonmagnetic impurities,
by which $T_{\rm c}$ is unchanged due to Anderson's theorem \cite{Schrieffer}.
In the Nambu representation \cite{Schrieffer}, 
a single impurity potential term is given by
\begin{eqnarray}
{\hat H}_{\rm imp} = \frac{I}{2} \sum_{\k,\k'} {\hat \phi}_\k^\dagger 
{\hat T}_0 {\hat \phi}_{\k'}, 
 \label{eqn:Himp}
\end{eqnarray}
where $I$ is the impurity potential.
Here and hereafter, we define the following $4\times4$ unitary matrices:
\begin{eqnarray}
{\hat T}_0 &=& 
\left(
\begin{array}[c]{cc}
{\hat 1} & 0 \\
0 & -{\hat 1}
\end{array}
\right), \ \
{\hat T}_1 = 
\left(
\begin{array}[c]{cc}
{\hat \s}_x & 0 \\
0 & -{\hat \s}_x
\end{array}
\right), 
 \nonumber \\
{\hat T}_2 &=& 
\left(
\begin{array}[c]{cc}
{\hat \s}_z & 0 \\
0 & {\hat \s}_z
\end{array}
\right), \ \
{\hat T}_3 = 
\left(
\begin{array}[c]{cc}
0 &{\hat \s}_z  \\
{\hat \s}_z & 0
\end{array}
\right) . \ \
\end{eqnarray}
The Green function in the Nambu representation is 
${\hat G}_\k(\w) = ( \w{\hat 1}-{\hat H}_\k )^{-1}$ \cite{Schrieffer},
and the retarded and advanced Green function are
${\hat G}_\k^R(\w)={\hat G}_\k(\w+i\gamma)$ and
${\hat G}_\k^A(\w)={\hat G}_\k(\w-i\gamma)$:
In the Born approximation, the damping rate is given by
$\gamma=n_{\rm imp}\pi I^2 N(\w)$, where
$N(\w)= (m/2\pi) {\rm Re}\{1/\sqrt{1-(\Delta/\w)^2}\}$
is the density of states (DOS) per spin.

Since the SHC is independent of the Meissner current, it will be 
insensitive to the electric field frequency $\w$ for $|\w|\ll \g_0$.
In the linear response theory \cite{Tanaka-4d5d,PALee},
the DC SHC is expressed in the $4\times4$ Nambu representation as
\begin{eqnarray}
\s_{\rm SH}^{I}&=&\sum_\k \int \frac{d\w}{2\pi}\left(-\frac{\d f}{\d \w}\right)
\frac12 {\rm Tr}\left[{\hat J}_x^{\rm S} {\hat G}^R {\hat \Lambda}_y^{\rm C} {\hat G}^A \right],
  \label{eqn:sxy-I}
 \\ 
\s_{\rm SH}^{I\!I}&=&-\sum_\k \int \frac{d\w}{2\pi}f(\w)
{\rm Re }\frac12{\rm Tr}\left[{\hat J}_x^{\rm S} \frac{\d{\hat G}^R}{\d\w}
{\hat J}_y^{\rm C} {\hat G}^R \right.
 \nonumber \\ 
& &\left. - {\hat J}_x^{\rm S} {\hat G}^R{\hat J}_y^{\rm C} \
\frac{\d{\hat G}^R}{\d\w}  \right],
  \label{eqn:sxy-II}
\end{eqnarray}
where $f(\w)=(e^{\w/T}+1)^{-1}$.
${\hat J}_y^{\rm C}$ is the bare charge current operator, which is given by 
$-\delta {\hat h}_{\k+e{\bf A}}/\delta A_y|_{{\bf A}=0}$
($\delta {\hat h}^*_{-\k+e{\bf A}}/\delta A_y|_{{\bf A}=0}$)
for particle (hole) channel \cite{PALee}.
${\hat J}_x^{\rm S} \equiv \{{\hat J}_x^{\rm C},s_z\}/(-2e)$
is the bare spin current operator.
In the present model, they are given by
\begin{eqnarray}
{\hat J}_y^{\rm C}= -e\left(\frac{k_y}{m}{\hat 1}+\lambda{\hat T}_1 \right), 
\ \ \
{\hat J}_x^{\rm S}= \frac{k_x}{2m}{\hat T}_2 ,
 \label{eqn:JCS}
\end{eqnarray}
where $-e\lambda{\hat T}_1 $ in ${\hat J}_y^{\rm C}$
is called the anomalous velocity \cite{Sinova-SHE,Inoue-SHE},
which is essential for the SHE.

${\hat \Lambda}_y^{\rm C}$ in eq. (\ref{eqn:sxy-I}) 
is the total charge current dressed by the CVC.
When the elastic scattering dominates the inelastic scattering,
the CVC is derived from the following Bethe-Salpeter equation in the 
self-consistent Born approximation (SCBA):
\begin{eqnarray}
{\hat \Lambda}_y^{\rm C} &=& {\hat J}_y^{\rm C} 
+ \Delta{\hat \Lambda}_y^{{\rm C}}, 
 \label{eqn:CVC}
\end{eqnarray}
The second term, 
$\Delta{\hat \Lambda}_y^{{\rm C}} \equiv 
n_{\rm imp}I^2 \sum_\k {\hat T}_0 {\hat G}^R {\hat \Lambda}_y^{\rm C} 
{\hat G}^A {\hat T}_0$,
represents the CVC for $I$-term. 
As shown in Ref. \cite{Inoue-SHE},
the factor $n_{\rm imp}$ in eq. (\ref{eqn:CVC}) cancels with 
$\sum_\k{\hat G}^R{\hat G}^A \sim O(\gamma^{-1})\sim O(n_{\rm imp}^{-1})$.
Therefore, the CVC for $I$-term is 
important even in the clean limit ($n_{\rm imp}\ll1$).
We will show this fact explicitly in later calculation.

On the other hand, CVC for $I\!I$-term is negligible in the clean limit:
The charge CVC for $I\!I$-term, $\Delta{\hat \Lambda}_y^{{\rm C}I\!I}$,
is given by eq. (\ref{eqn:CVC}) by replacing ${\hat G}^A$ with ${\hat G}^R$.
Since $\sum_\k{\hat G}^R{\hat G}^R$ is regular for $\gamma\rightarrow+0$,
$\Delta{\hat \Lambda}_y^{{\rm C}I\!I} \sim O(n_{\rm imp})$,
which is negligible in the clean limit.
Moreover,
$\Delta{\hat \Lambda}_y^{{\rm C}I\!I}$ is related to the self-energy 
${\hat \Sigma}_\k$ through the Ward identity.
Since ${\hat \Sigma}_\k$ is $\k$-independent in the SCBA 
for short-range impurities,
$\Delta{\hat \Lambda}_y^{{\rm C}I\!I}$ vanishes 
in the present study.
Thus, bare currents enter into eq. (\ref{eqn:sxy-II}).

By following Ref. \cite{Tanaka-4d5d}, 
we rewrite eqs. (\ref{eqn:sxy-I}) and (\ref{eqn:sxy-II})
in the band-diagonal basis.
In this basis, $({\hat U}_\k^\dagger {\hat G}^R {\hat U}_\k)_{l,m}
\equiv[{\hat G}^R]_{l,m}=\delta_{l,m}/(\w-E_\k^{l})$,
where $E_\k^i= \pm E_{\k,\pm}$ and 
${\hat U}_\k$ is the unitary matrix for the change of basis.
Then, we can perform $\w$-integrations using the relationships
${\rm Im}(z-i\gamma)^{-1}=\pi \delta(z)$ and
${\rm Im}(z-i\gamma)^{-2}=-\pi \delta'(z)$.
For example, $\s_{\rm SH}^{I\!I}$ is given as
%
\begin{eqnarray}
\frac12 \sum_{\k,l\ne m} 
\frac{[J\!J]_{m,l}}{E_\k^m-E_\k^l}
\left[ \left(\frac{\d f}{\d \w}\right)_{E_\k^l} 
+2f(E_\k^l)\frac1{E_\k^m-E_\k^l} \right] ,
 \nonumber 
\end{eqnarray}
where $[J\!J]_{m,l}\equiv {\rm Im}\{[{\hat J}_x^{\rm S}]_{m,l}[{\hat J}_y^{\rm C}]_{l,m}\}$.
The first (second) term corresponds to 
$\s_{\rm SH}^{I\!Ia}$ ($\s_{\rm SH}^{I\!Ib}$) in Refs. \cite{Tanaka-4d5d}.
As a result, $\s_{\rm SH}$ is given by the summation of 
$\s_{\rm SH}^{I+I\!Ia}$ and $\s_{\rm SH}^{I\!Ib}$:
%
\begin{eqnarray}
\s_{\rm SH}^{I+I\!Ia} 
&=&\sum_{\k,l\ne m} 
\frac{[J\!\Delta \Lambda]_{m,l}}{2(E_\k^m-E_\k^l)}
\left(-\frac{\d f}{\d \w}\right)_{E_\k^l} ,
  \label{eqn:I-IIa} \\ 
\s_{\rm SH}^{I\!Ib}&=&\sum_{\k,l\ne m} 
\frac{[J\!J]_{m,l}}{2(E_\k^m-E_\k^l)^2}
\left[ f(E_\k^l)-f(E_\k^m) \right] ,
 \label{eqn:II-b} 
\end{eqnarray}
where $[J\!\Delta \Lambda]_{m,l}\equiv {\rm Im}\{[J_x^{\rm S}]_{m,l}[\Delta\Lambda_y^{\rm C}]_{l,m}\}$.
Now, we can calculate the SHC in general superconductors
using Eqs. (\ref{eqn:I-IIa}) and (\ref{eqn:II-b}), 
which had not been derived previously.

\begin{figure}[ptbh]
\includegraphics[width=.5\linewidth]{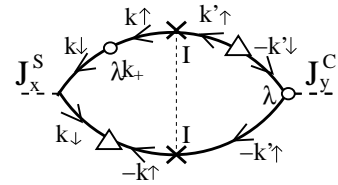}
\caption{An example of terms for $\s_{\rm SH}^I$
that is finite under $T_{\rm c}$.
This term is proportional to $\Delta \propto \sqrt{1-T/T_{\rm c}}$.
}
\label{fig:Diagram}
\end{figure}

$\s_{\rm SH}^{I+I\!Ia}$ in eq. (\ref{eqn:I-IIa})
represents the quasiparticle contribution at the Fermi level,
and $\s_{\rm SH}^{I\!Ib}$ in eq. (\ref{eqn:II-b})
represents the Berry curvature term that is rewritten as the summation of 
the Berry curvature of Bloch electrons inside the Fermi sea.
In the normal state in bulk transition metals,
CVC due to local impurities almost vanishes in the clean limit
since the main matrix elements of current operators are odd functions of $\k$.
Therefore, $\s_{\rm SH}^{I+I\!Ia}\approx0$ and the total SHC is approximately 
given by $\s_{\rm SH}^{I\!Ib}$ for $\gamma\rightarrow0$ \cite{Tanaka-4d5d}.
In the Rashba 2DEG model, in contrast,
$\s_{\rm SH}^{I+I\!Ia}=e/8\pi$ in the normal state 
because of the CVC for $I$-term.
Since $\s_{\rm SH}^{I\!Ib}=-e/8\pi$,
the total SHC vanishes identically in the normal state \cite{Inoue-SHE}.

Interestingly, such a balance between the quasiparticle term
and the Berry curvature term in the normal state is drastically changed
when the superconductivity ($\Delta\ne0$) sets in:
$\s_{\rm SH}^{I+I\!Ia}$ vanishes at $T=0$ because of the factor $-\d f/\d\w$.
Thus, $\s_{\rm SH}^{I+I\!Ia}$ should show a prominent change below 
$T_{\rm c}$ in spin Hall systems
where the CVC is significant:
such examples are metals with Rashba-type or Dresselhaus-type SOI
\cite{Inoue-SHE,Nomura-SHE} 
and doped graphene \cite{graphene}.
On the other hand, $\s_{\rm SH}^{I\!Ib}$ is finite even at $T=0$ 
unless $\Delta$ is much larger than the SOI splitting.
Therefore, the SHC can show a nontrivial temperature dependence 
below $T_{\rm c}$ due to the imbalance between 
$\s_{\rm SH}^{I+I\!Ia}$ and $\s_{\rm SH}^{I\!Ib}$.

Hereafter, we calculate both $\s_{\rm SH}^{I+I\!Ia}$ and 
$\s_{\rm SH}^{I\!Ib}$ in the BCS-Rashba 2DEG model,
and show that the SHC takes a large value just below $T_{\rm c}$.
According to eq. (\ref{eqn:CVC}), 
the lowest order CVC for ${\Delta\Lambda_y^{\rm C}}$ is given as
\begin{eqnarray}
{\Delta{\hat \Lambda}_y^{\rm C}}^{(1)}(\w)&=&
n_{\rm imp}I^2 \sum_\k {\hat A}(\w) ,
 \label{eqn:CVC1} 
\end{eqnarray}
where 
${\hat A}(\w)
= {\hat T}_0 {\hat G}^R {\hat J}_y^{\rm C} {\hat G}^A {\hat T}_0$.
Hereafter, we assume the damping rate in the normal state,
$\gamma_0= n_{\rm imp}I^2 m/2$, is the smallest parameter.
Considering that the pole of ${\hat G}^R$ is $\w=\pm E_\pm -i\gamma$
and the relation
$(z^2+\gamma^2)^{-1}=(\pi/\gamma)\delta(z)$,
the expression of ${\hat A}(\w)$ is greatly simplified for $\gamma_0\ll1$.
After performing the integration $\int_0^{2\pi}d\phi_k/2\pi$
($\phi_k= {\rm tan}^{-1}(k_y/k_x)$), ${\hat A}(\w)$ is simply given as
\begin{eqnarray}
{\hat A}(\w)= \frac{-e\cdot\pi}{8\gamma} \sum_{\a=\pm1}
 \left(\a \frac{k}{m}+\lambda\right)
\delta(|\w|-E_\a)\left({\hat T}_1 -x{\hat T}_3 \right)
 \nonumber 
\end{eqnarray}
where $x=\Delta/\w$.
As a result, eq. (\ref{eqn:CVC1}) becomes
\begin{eqnarray}
{\Delta{\hat \Lambda}_y^{\rm C}}^{(1)}(\w)&=& -e(\lambda/2)
(-{\hat T}_1+ x{\hat T}_3) ,
\end{eqnarray}
where the relation $\gamma= \pi n_{\rm imp}I^2N(\w)$ is considered.

The second lowest order CVC
is given by eq. (\ref{eqn:CVC1}) by replacing ${\hat J}_y^{\rm C}$
with ${\Delta{\hat \Lambda}_y^{\rm C}}^{(1)}$.
After taking the $\k$-summation, it is simply obtained as
\begin{eqnarray}
{\Delta{\hat \Lambda}_y^{\rm C}}^{(2)}(\w) 
&=& (1-x^2)/2 \cdot {\Delta{\hat \Lambda}_y^{\rm C}}^{(1)}(\w) .
\end{eqnarray}
Thus, the total CVC,
$\Delta\Lambda_y^{\rm C}= \sum_{l=1}^\infty{\Delta{\hat \Lambda}_y^{\rm C}}^{(l)}$,
is given as
\begin{eqnarray}
\Delta\Lambda_y^{\rm C}
&=& -e\frac{\lambda}{1+x^2}\left( -{\hat T}_1 +x {\hat T}_3 \right) .
 \label{eqn:Lam}
\end{eqnarray}
$\s_{\rm SH}^I$ vanishes in the normal state since
the off-diagonal anomalous velocity cancels out
in the total charge current for $x=0$; $\Lambda_y^{\rm C}= J_y^{\rm C}+
\Delta\Lambda_y^{\rm C} = -e(k_y/m){\hat 1}$ \cite{Inoue-SHE}.
However, this cancellation does not occur in the superconducting state,
and therefore $\s_{\rm SH}^I\ne0$ below $T_{\rm c}$.

Now, we derive $\s_{\rm SH}^{I+I\!Ia}$
that represents the quasiparticle contribution.
In a general basis, eq. (\ref{eqn:I-IIa}) is rewritten as
\begin{eqnarray}
\s_{\rm SH}^{I+I\!Ia} 
= \sum_\k \int \frac{d\w}{2\pi} \left(-\frac{\d f}{\d\w}\right) B(\w) 
  \label{eqn:I-IIa-2}
\end{eqnarray}
where $B(\w)\equiv \frac12 {\rm Tr}\left[{\hat J}_x^{\rm S} 
{\hat G}^R \Delta{\hat \Lambda}_y^{\rm C} {\hat G}^A \right]$.
Since $B(\w)$ is simplified as
$(e/32m(1+x^2))\sum_{\a=\pm1} (\e_\k+\a\lambda k)^2\delta(|\w|-E_\a)$
for $\gamma\ll1$, we obtain the following expression:
\begin{eqnarray}
\s_{\rm SH}^{I+I\!Ia} &=& \frac{e}{8\pi}X(\Delta,T) ,
 \label{eqn:I-IIa-res}
 \\
X(\Delta,T)
&=& \int_{-\infty}^\infty d\e 
\left(-\frac{\d f}{\d\w}\right)_{\sqrt{\e^2+\Delta^2}}
\frac{\e^2}{\e^2+2\Delta^2} .
 \nonumber 
\end{eqnarray}
It should be stressed that
$\s_{\rm SH}^{I+I\!Ia}$ is independent of $\lambda(\ne0)$.
The asymptotic behavior of $X(\Delta,T)$ for $\Delta\ge0$ is
\begin{eqnarray}
X(\Delta,T)&\approx& 1-\frac{\pi}{2\sqrt{2}}\frac{\Delta}{T_{\rm c}}
\ \ \ \mbox{for $\Delta\ll T$ ($T\sim T_{\rm c}$)} ,
 \\
&\approx& \sqrt{\frac{\pi T}{2\Delta}}e^{-\Delta/T}
\ \ \mbox{for $\Delta\gg T$ ($T\ll T_{\rm c}$)} .
\end{eqnarray}
In the same way, we derive that 
$\s_{\rm SH}^{I\!Ia}= (e/8\pi)Y(\Delta,T)$, where
$Y=\int_{-\infty}^\infty d\e 
(-\d f/\d\w)_{\sqrt{\e^2+\Delta^2}}$ is the Yosida function.
Since $Y=1-O((\Delta/T_{\rm c})^2)$ for $T\sim T_{\rm c}$,
$\s_{\rm SH}^I \approx (-e/16\sqrt{2})(\Delta/T)$.
Figure \ref{fig:Diagram} shows an example of terms for $\s_{\rm SH}^{I}$
that is of order $O(\Delta)$ after $\k,\w$-integrations.
It represents the spin current due to the triplet 
$(\k \uparrow,-\k \uparrow)$ particle-particle excitation
induced by the Rashba SOI and the impurity scattering.

Next, we discuss $\s_{\rm SH}^{I\!Ib}$, which is called the 
``Berry curvature term'' \cite{Tanaka-4d5d}.
It is caused by electrons 
in the Fermi sea that satisfy $E_k^l\cdot E_\k^m<0$.
In the normal state, the $\k$-summation in eq. (\ref{eqn:II-b})
is restricted to $k_{\rm F-}<|\k|<k_{\rm F+}$, where $k_{\rm F\pm}$ are 
two Fermi momenta \cite{Tanaka-4d5d}.
This restriction approximately holds in the superconducting state in the present model.
Moreover, $[{\hat J}_x^{\rm S}]_{m,l}=[{\hat J}_x^{\rm C}]_{m,l}=0$ for 
$(E_\k^m,E_\k^l)=(E_{\k,\a},-E_{\k,\a})$,
since there is no number-nonconserving elements in
eq. (\ref{eqn:JCS}) \cite{PALee}.
For these reasons, in contrast to $\s_{\rm SH}^{I+I\!Ia}$, 
$\s_{\rm SH}^{I\!Ib}$ is insensitive to 
$T$ and $\Delta$ if $(T, \Delta) \ll \Delta_{\rm SO}$,
which will be verified later by performing numerical 
calculation of eq. (\ref{eqn:II-b}).

\begin{figure}[ptbh]
\includegraphics[width=.85\linewidth]{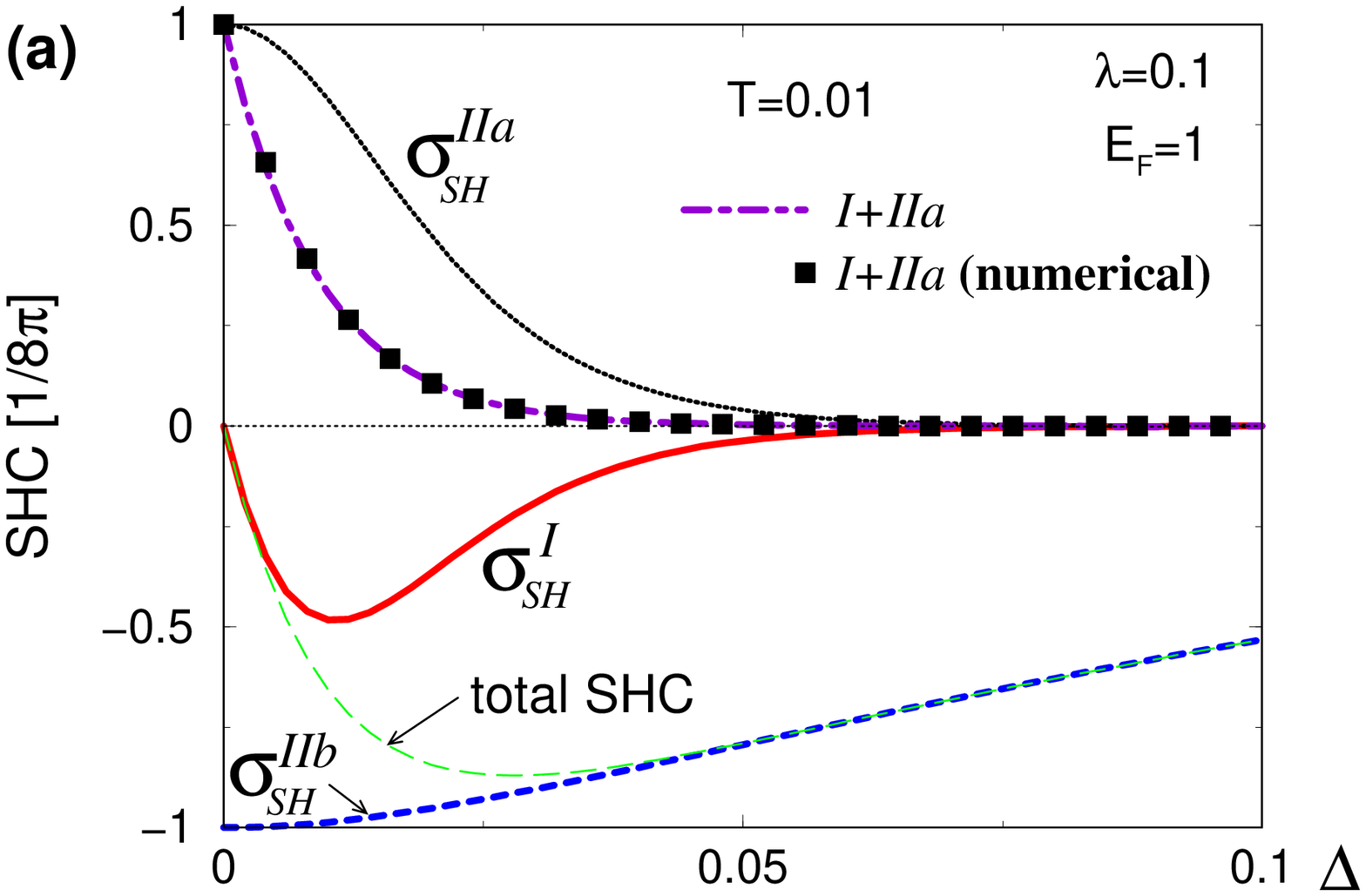}
\includegraphics[width=.85\linewidth]{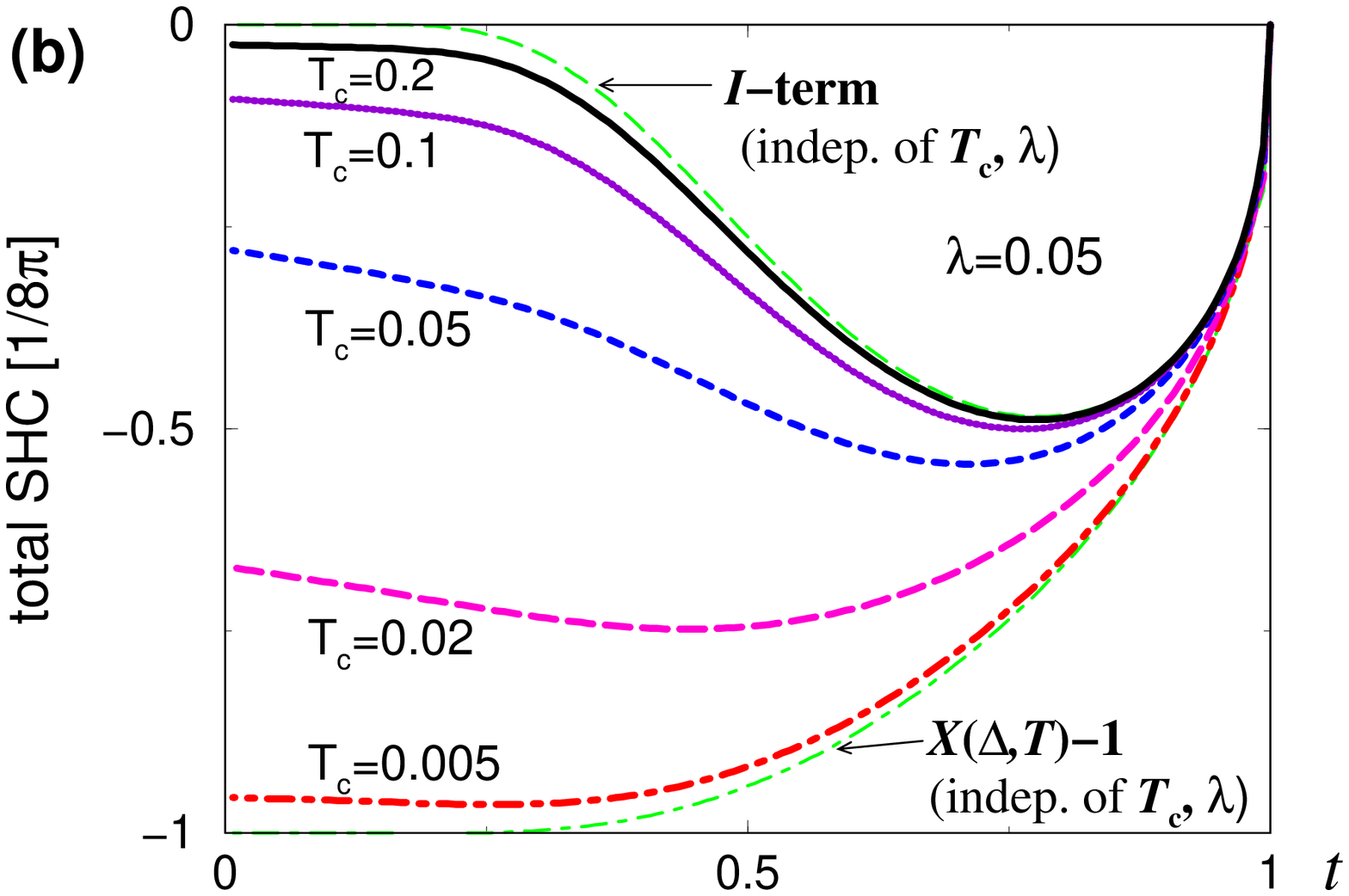}
\caption{(color online)
{\bf (a)} $\Delta$-dependence of the SHC; 
comparison of $I$-, $I\!Ia$-, and $I\!Ib$-terms.
{\bf (b)} $t\equiv T/T_{\rm c}$-dependence of the total SHC, $\s_{\rm SH}$,
for 
$T_{\rm c}=0.2\sim0.005$.
}
\label{fig:FIG}
\end{figure}

Now, we show numerical results for $m=\mu=1$,
under the condition that 
$\Delta,\Delta_{\rm SO} \gg \gamma_0$.
Figure \ref{fig:FIG} (a) shows the $\Delta$-dependence of the SHC
for $\lambda=0.1$ and $T=0.01$.
The quasiparticle term $\s_{\rm SH}^{I+I\!Ia}$ and 
the Berry curvature term $\s_{\rm SH}^{I\!Ib}$ are given by
eqs. (\ref{eqn:I-IIa-res}) and (\ref{eqn:II-b}), respectively.
To verify the correctness of eq. (\ref{eqn:I-IIa-res}),
we have also calculated the CVC by solving
the Bethe-Salpeter eq. (\ref{eqn:CVC}) numerically
for $\gamma_0=0.001$, and derived
$\s_{\rm SH}^{I+I\!Ia}$ using eq. (\ref{eqn:I-IIa-2}).
The obtained result is shown by square dots.
Since $\s_{\rm SH}^{I}= -(e/16\sqrt{2})(\Delta/T_{\rm c}) + O(\Delta^2)$
and $\s_{\rm SH}^{I\!I}= O(\Delta^2)$,
$\s_{\rm SH}^{I}$ gives the increment in 
$|\s_{\rm SH}|$ just below $T_{\rm c}$.

Figure \ref{fig:FIG} (b) shows the $T$-dependence of the SHC
for $\lambda=0.05$, assuming that $\Delta=\Delta_0\sqrt{1-T/T_{\rm c}}$
and $\Delta_0=1.8 T_{\rm c}$ in accord with the weak-coupling BCS theory.
It is noteworthy that 
both $\s_{\rm SH}^{I}$ and $\s_{\rm SH}^{I+I\!Ia}$
are unique functions of $t\equiv T/T_{\rm c}$, and independent of
$T_{\rm c}$ and $\Delta_{\rm SO}$.
$\s_{\rm SH}^{I}$ takes a minimum value 
$\s_{\rm SH}^{I {\rm max}}\approx -0.48$ at $t\approx 0.78$.
In the case of $\Delta_0 \ll \Delta_{\rm SO}$
(i.e., $T_{\rm c} \ll \Delta_{\rm SO}$),
$\s_{\rm SH}^{I\!Ib} \approx -e/8\pi$ for $0<T<T_{\rm c}$
since the modification of electronic states for $k_{\rm F-}<|\k|<k_{\rm F+}$
is limited, as discussed above.
For this reason, $\s_{\rm SH} \approx -e/8\pi\cdot(X(\Delta,T)-1)$ 
for $\Delta_0 \ll \Delta_{\rm SO}$, as shown in Fig. \ref{fig:FIG} (b). 
$\s_{\rm SH}(T=0)=\s_{\rm SH}^{I\!Ib}(T=0)$ decreases with $\Delta_0$ 
since the formation of singlet paring prevents the spin current.
In the opposite case, $\Delta_0 \gg \Delta_{\rm SO}$,
the relation $\s_{\rm SH} \sim \s_{\rm SH}^I$ is recognized,
which means that
$\s_{\rm SH}^{I\!I}=\s_{\rm SH}^{I\!Ia}+\s_{\rm SH}^{I\!Ib}$ is very small.
Thus, SHC shows a striking enhancement just below $T_{\rm c}$
for any value of $\lambda$.

In the present study, we consider that the conduction electrons are 
delocalized for $k_{\rm F-}<|\k|<k_{\rm F+}$, and they contribute to 
$\s_{\rm SH}^{I\!Ib}$.
In conventional insulators, in contrast, $\s_{\rm SH}^{I\!Ib}$
is expected to vanish since the electrons near the band edge
are localized \cite{Kane,SCZhang}. 
(Note that localization cannot be described in the SCBA.)
Interestingly, recent theoretical and experimental efforts have 
revealed that $\s_{\rm SH}^{I\!Ib}$ takes a finite value
in ``topological insulators'' such as graphene \cite{Kane} 
and HgTe \cite{SCZhang},
owing to the delocalized nature of massless Dirac fermions.

We have shows that the SHE driven by $k$-linear 
Rashba SOI shows a prominent increment in the superconducting state.
Similar drastic change in the SHC will be observed in cases of the
$k^3$-type Rashba or Dresselhaus SOI, since the CVC for the 
anomalous velocity is large in these cases.
The present study opens the way to 
distinguish between the quasiparticle contribution (eq. (\ref{eqn:I-IIa}))
and the Berry curvature contribution (eq. (\ref{eqn:II-b})),
which had been desired to understand the 
mechanism of intrinsic SHE.


In summary, we have presented the first study of the
intrinsic SHE in the superconducting state.
In the Rashba 2DEG model, the SHC changes from zero
to a large negative value just below $T_{\rm c}$
unless $\lambda=0$, due to the change in the
quasiparticle contribution $\s_{\rm SH}^{I+I\!Ia}$.
This phenomenon will be observed in thin film superconductors
with the aid of surface-induced Rashba SOI,
in non-centrosymmetric superconductors,
and in semiconductors using the superconducting proximity effect.
Moreover, the SHC in superconductors
will also be observed using the AC measurement, 
since $\s_{xx}(\w)$ is finite for $\omega\ne0$ whereas
the intrinsic SHC is independent of $\omega$ for $0\le\w\lesssim \gamma^{-1}$
\cite{Inoue-SHE}.

We are grateful to Y.Otani, T. Kimura, and E. Saitoh 
for valuable discussions.


\end{document}